\documentclass{aa}  

\usepackage{graphicx}
\usepackage{txfonts}
\begin{document}

   \title{Chromatic line-profile tomography to reveal exoplanetary atmospheres: application to \object{HD~189733b}}
   \titlerunning{Chromatic line-profile tomography of HD~189733b}
   \authorrunning{F.~Borsa et al.}

   \author{F.~Borsa
          \inst{1},
          M.~Rainer\inst{1}
          \and
          E.~Poretti\inst{1}
          }

   \institute{INAF -- Osservatorio Astronomico di Brera, Via E. Bianchi 46, 23807 Merate (LC), Italy\\
              \email{francesco.borsa@brera.inaf.it}
             }

   \date{Received ; accepted }

 
  \abstract
   {Transmission spectroscopy can be used to constrain the properties of exoplanetary atmospheres. During a transit, the light blocked from the atmosphere of the planet leaves an imprint in the light coming from the star. This has been shown for many exoplanets with different techniques, with both photometry and spectroscopy.}
   {We aim at testing chromatic line-profile tomography as a new tool to investigate exoplanetary atmospheres. The signal imprinted on the cross-correlation function (CCF) by a planet transiting its star is dependent on the planet-to-star radius ratio. We want to verify if the precision reachable on the CCF obtained from a subset of the spectral orders 
   of the HARPS spectrograph is enough to discriminate the radius of a planet at different wavelengths. }
   {We analyze HARPS archival data of three transits of HD~189733b. 
   We divide the HARPS spectral range in 7 broadbands, calculating for each band the ratio between the area of the out-of-transit CCF and the area of the signal imprinted by the planet on it during the full part of the transit. We take into account the effect of the limb darkening using the theoretical coefficients of a linear law.
   Averaging the results of three different transits allows us to obtain a good quality broadband transmission spectrum of HD~189733b, with a precision greater than that of the chromatic Rossiter McLaughlin effect. 
   }
   {We proved chromatic line-profile tomography to be an interesting way to reveal broadband transmission spectra of exoplanets:
   our analysis of the atmosphere of HD~189733b is in agreement with other ground- and space-based observations.
   The independent analysis of different transits puts in evidence the probability that stellar activity plays a role in the extracted transmission spectrum. Care has thus
   to be taken when claiming for Rayleigh scattering in the atmosphere of exoplanets orbiting active stars using only one transit.}
   {}


   \maketitle
%

\section{Introduction\label{sec:intro}}

The number of discovered exoplanets is growing exponentially in the last years, thanks to a lot of ground- and space-based surveys, taking advantage of different methods of detection \citep[e.g.,][]{wrightgaudi}. While the existence of exoplanets, and their abundance in our galaxy, is now a firmly established knowledge, the attention of the scientific community is moving to exoplanet characterization and atmospheric content.
During a planetary transit, a fraction of the light coming from the host star is filtered through the planet's atmosphere, thus providing hints of its composition.
The atmospheric depth at which the stellar light becomes extinct depends on the wavelength and atmospheric composition, and if an atom or molecule is present, the planet becomes opaque in the respective absorption band, resulting in a larger obscured area. As a consequence, the eclipse appears deeper at the wavelength of the absorption band, yielding an apparent larger radius.
The study of this effect, i. e. transmission spectroscopy, has been demonstrated to be
a very successful technique for investigating exoplanetary atmospheres \citep[][]{burrows}.

HD~189733b \citep{bouchy} represents the perfect science case for testing new methods aimed at identifying its transmission spectrum. 
The brightness of its host star ($M_V=7.67$) and the depth of its transit (>2\%) permit to reach a high signal-to-noise ratio (S/N).
Its atmosphere at optical wavelengths has been explored with many different methods using primary transit, with both ground- and space-based observations.
From Earth, \citet{redfield} were the first to detect sodium absorption using the High Resolution Spectrograph on the Hobby-Eberly telescope. 
Later, \citet{wyttenbach} could evidence the signal of atomic sodium 
and \citet{louden} could retrieve the velocity of winds on both the hemispheres of the planet with HARPS.
With the same observations \citet{digloria} probed the Rayleigh scattering slope by using the chromatic Rossiter-McLaughlin (RM) effect.
\citet{McCullough} claimed that stellar activity is the main responsible for the slope observed in the transmission spectrum.
From space, \citet{pontACS} with ACS and \citet{singSTIS} with STIS detected Rayleigh scattering using the {\it Hubble Space Telescope} (HST). 
\citet{pont} presented a homogeneous reanalysis of the data and a global transmission spectrum from ultraviolet to infrared.
\citet{huitson} could detect and characterize sodium in the atmosphere.

\

Each surface element of a rotating star has a specific radial velocity. 
If there is a reduction of flux from a region, for example for the presence of a spot or for the obscuration by a transiting planet, the contribution to the stellar line-profile at this velocity will be affected, resulting in the presence of a {\it bump} in the line-profile.
As the transiting planet moves across the disk, this perturbation will move across the line-profile, because the radial velocity of the obscured zone is different.
The analysis of this effect, called line-profile tomography, has been very useful for detecting the inclination of the orbital plane of the planet with respect to the stellar rotational axis \citep[e.g.,][]{cameron,cameron2}, for the confirmation of the presence of exoplanets \citep[][]{hartman}, for the measurement of their nodal precession \citep[][]{johnson}. This has been done in particular for high $V_{\rm eq}\sin{i}$ stars, because for these stars it is more difficult to extract precise radial velocities and analyze the RM effect.
Line-profile variation studies have been also widely used in the study of the stars themselves, with asteroseismology. The spectral lines of a pulsating variable show variable mean line-profiles. 
By measuring the observed variations, it is possible to typify the excited modes by assigning the spherical wavenumbers ({\it l},m) to each of them \citep[e.g.,][]{poretti}.


To compare the atmospheric properties between different exoplanets and subsequently identifying possible sub-classes, we need to observe as more exoplanetary atmospheres as possible. Given the difficulty of this kind of observations, any new method of investigation is welcome to increase the sample.
In this paper, we will describe the chromatic line-profile tomography as a new method for obtaining broadband transmission spectroscopy of exoplanetary atmospheres. 
We will test it by using the well known case of the transiting exoplanet HD~189733b and comparing the results with its transmission spectra obtained from other analyses.


\section{HARPS data and choice of broadbands\label{sec:data}}

As in previous atmospheric analyses of ground-based data of HD~189733b, we used the three full transits observed spectroscopically with HARPS, available in the ESO archive and obtained under programs 072.C-0488(E), 079.C-0127(A) and 079.C-0828(A). An observations log, together with the number of exposures and exposure times for each transit, is available in \citet[][see their Table~1]{wyttenbach}.

HARPS \citep{mayor} is a fiber-fed echelle spectrograph composed of 72 orders, in a range of wavelength from 378 nm to 691 nm and a resolving power $R\simeq115000$.
The HARPS Data Reduction Software (DRS) estimates the radial velocities (RVs) of the targets by computing cross-correlation functions \citep[CCFs, ][]{baranne,2002A&A...388..632P} between a weighted line mask and the stellar spectrum for each order and then summing them together. 
To avoid contamination from the telluric lines, the masks do not have any reference line in the wavelength ranges where the terrestrial atmospheric contribute is very large. Because of this, there is no CCF at all for three orders of the spectra (orders 89, 94 and 103).


The method we are testing here uses CCFs obtained from different subsets of orders. We chose to use the same orders division as used by \citet{digloria} in order to better check our results and the reliability of our method (see Table~\ref{tabLD}).
After a first analysis, we considered that the use of the first 9 orders (up to the \ion{Ca}{ii} H{\&}K spectral region) could led to unreliable results.
The \ion{Ca}{ii} H{\&}K lines, in fact, are particularly variable in the case of active stars \citep[such as HD~189733,][]{boisse}, and are often used as a tracer for the level of stellar activity \citep[e.g.,][]{borsa}. The emission present in their core could thus influence the shape of the CCF. Moreover, the S/N of these orders is by far lower than in the others for the star analyzed: this is due to the low efficiency of the instrument at these wavelengths and to the spectral type of the star \citep[K1-K2,][]{bouchy}.
Stellar and transit parameters used during this analysis are listed in Table~\ref{tabParameters}.

 \begin{table}
\begin{center}
\caption{Wavelength ranges for the order groups considered, and the respective linear limb darkening coefficient used. The last line refers to the total HARPS wavelength range.}
\label{tabLD}
\footnotesize
\begin{tabular}{cccc}
 \hline\hline
 \noalign{\smallskip}
Group & Passband (nm) & Orders & $u$\\
\noalign{\smallskip}
\hline
\noalign{\smallskip}
1 & 400-420  & 146-152 & $0.9332\pm0.0009$\\
2 & 420-470  & 131-145 & $0.9043\pm0.0015$\\
3 & 470-520  & 118-130 & $0.8548\pm0.0015$\\
4 & 520-570  & 107-117 & $0.7917\pm0.0017$\\
5 & 550-600  & 104-110 & $0.7625\pm0.0018$\\
6 & 600-650  & 95-101 & $0.7073\pm0.0017$\\
7 & 650-700  & 90-92 & $0.6596\pm0.0017$\\
HARPS & 370-700  & 90-161 & $0.7962\pm0.0013$\\
\noalign{\smallskip}
 \hline
\end{tabular}
\end{center}
\end{table}

 \begin{table}
\begin{center}
\caption{Stellar and transit parameters used in this work.}
\label{tabParameters}
\footnotesize
\begin{tabular}{ccc}
 \hline\hline
 \noalign{\smallskip}
 Parameter & Value & Reference\\
 \noalign{\smallskip}
\hline
\noalign{\smallskip}
\multicolumn{3}{c}{\it Stellar parameters}\\
\noalign{\smallskip}
T [K]&   $4875\pm43$ & \citet{boyajian}\\
log g &   $4.56\pm0.03$ & \citet{boyajian}\\
 Fe/H &   $-0.03\pm0.08$ & \citet{torres}\\
\noalign{\smallskip}
\multicolumn{3}{c}{\it Transit parameters}\\
\noalign{\smallskip}
Period [days]&   $2.21857312$ & \citet{triaud}\\
$T_0$ [BJD] &   $2453988.80339$ & \citet{triaud}\\  
$R_{\rm p}$/$R_{\rm s}$ &   $0.1581$ &  \citet{triaud}\\
$R_{\rm s}/a$ &   $0.1142$ & \citet{triaud}\\
$e$ &   $0.0$ & assumed\\
$i$ [degrees]&   $85.508$ & \citet{triaud}\\
\noalign{\smallskip}
 \hline
\end{tabular}
\end{center}
\end{table}


\section{Chromatic line-profile tomography\label{sec:shadow}}

The morphology of the CCF changes when a planet transits in front of the star: the Gaussian profile is no more symmetric, and this asymmetry causes 
a systematic deviation when measuring the velocity centroid of the starlight. This effect -- the Rossiter-McLaughlin effect -- causes a spurious deviation of the RV measurements from the pure Keplerian motion, which is known to be an artifact of the asymmetry just described.

In the CCF, the missing light during the full part of the transit of a planet can be expressed with the formula \citep[Eq. 9 in ][]{cameron}:

\begin{equation}
\beta= \frac{R_{\rm p}^{2}}{R_{\rm s}^{2}} \frac{1-u+u\mu}{1-u/3}
\label{equazbeta}
\end{equation}
where $\mu=\sqrt{1-d^{2}}$, with $d$ the distance of the planet from the center of the star in stellar radius units, and $u$ the coefficient of linear limb darkening.
We refer the reader to \citet{cameron} for details, but what we emphasize here is the fact that $\beta$ depends on the planet-to-star radius ratio ($R_{\rm p}$/$R_{\rm s}$).

The limb darkening (LD) coefficients for the wavelength ranges of our chosen order groups, reported in Table~\ref{tabLD}, were calculated using the code \texttt{LDTK} \citep{hannu}, that permits to generate custom LD coefficients and the relative uncertainties with a library of \texttt{PHOENIX}-generated specific intensity spectra \citep{husser}. 
We assumed here that both the mean line-profile of the star and the shadow caused by the planet could be modeled with a Gaussian function.

\subsection{Data analysis\label{sec:description}}

We compute a separate analysis for each transit, in order to be free from systematics due to different nights of observations which can perturb the CCF shape (e.g., terrestrial atmospheric conditions).
One more aspect that favors this approach is that in this way we can compare the transmission spectra obtained on different nights, investigating
the possibility that stellar activity can influence the retrieval of the transmission spectrum. The Rayleigh scattering claimed in the planet's atmosphere has in fact been questioned to be just an artifact caused by stellar activity \citep{McCullough}.

As a first step, for each spectrum we compute the predicted RV of the host star relative to the pure Keplerian motion, without the apparent deviation caused by the RM. To do this we compute a linear fit using the DRS-estimated RVs of the out-of-transit spectra, and use this line as our reference for the Keplerian motion of the star caused by the planet.

\begin{figure}[!ht]
\centering
\includegraphics[width=\linewidth]{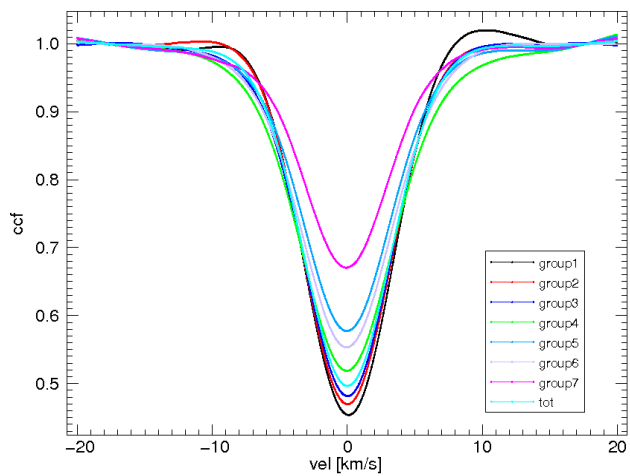}
\caption{Reference CCFs for the different groups of orders for the night Aug 28th, 2007.}
\label{fig:ccf_normalizzate}
\end{figure}

Then we apply our new procedure for each broadband chosen in Table~\ref{tabLD}.
We extract the CCFs relative to the orders of the broadband we are analyzing from the CCF fits files generated by the HARPS DRS pipeline.
These CCFs are then averaged and normalized, thus creating one CCF per spectrum for each broadband. The normalization is done in the same way for all the CCFs: with a linear fit between all the points with ${\rm v}<-15$ km\,s$^{-1}$ or ${\rm v}>15$ km\,s$^{-1}$, thus including in the fit only the continuum. We also changed the limit value to 13~km\,s$^{-1}$ and 17~km\,s$^{-1}$, without significant differences in the final results.
When speaking of CCF here afterwards in this section, we refer to these CCFs relative to the group of orders.
We create the reference CCF for the star (CCF$_{\rm ref}$) by averaging all the CCFs of the out-of-transit spectra (Fig.~\ref{fig:ccf_normalizzate}), shifted for the relative RVs (as estimated from the pure Keplerian motion). No significant offset among different groups was detected (Fig.~\ref{fig:ccf_normalizzate}).
We fit CCF$_{\rm ref}$ with a Gaussian function using the \texttt{IDL} function \texttt{GAUSSFIT}, and subsequently we calculate its area 
as $A_{\rm ref}=\sqrt{2\pi} \cdot a \cdot c$, where $a$ and $c$ are the height and the standard deviation of the Gaussian respectively. 
The Spearman test did not detect any significant correlation between them. Hence,
we also give the error-bars for the area by propagating the errors on the $a$ and $c$ parameters obtained from the fits.

\begin{figure*}[!ht]
\centering
\includegraphics[width=\linewidth]{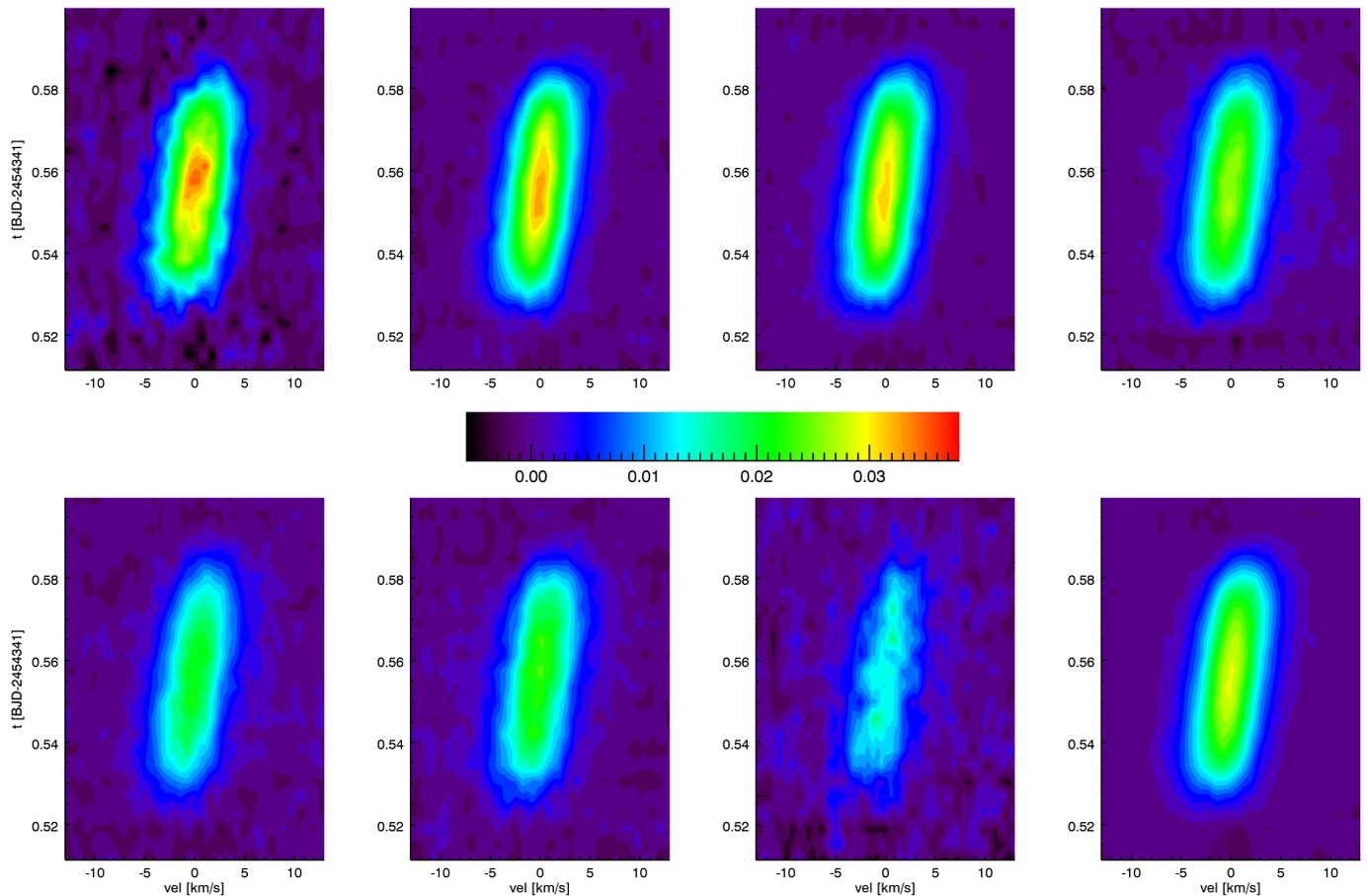}
\caption{
Shadow of the planet over the stellar CCF for the night Aug 28th, 2007 in the different order groups (Table~\ref{tabLD}) and in the total HARPS passband.
{\it Top:} left to right, shadow for the groups 1 to 4. {\it Bottom:} left to right, shadow for the groups 5 to 7, plus the total HARPS passband.
The color scale, chosen for viewing ease, is the same in all the plots. 
}
\label{fig:tomography}
\end{figure*}

The CCF$_{\rm ref}$ is then subtracted from all the CCFs of every exposure. 
A special care was taken while normalizing the data inside the transit: the decreasing of the photometric flux was taken into account, using a transit model created with the formalism of \citet{mandelagol} and the transit parameters of Table~\ref{tabParameters}.
The data are now ready to show the typical Doppler shadow of the planet passing in front of the star (Fig.~\ref{fig:tomography}).

From now on we will use only the data that are inside the full part of the transit.
On each of the residual data we fit a Gaussian profile, estimating its area in the same way as done for $A_{\rm ref}$. We paid particular attention in checking that the Gaussian profile was centered where the signal of the planet was expected.  
Given the fact that this signal is very low, in particular when using a small number of orders, this is important to avoid that the Gaussian profile we are fitting is centered on a random noise bump instead of on the planetary one.

\begin{figure}[!h]
\centering
\includegraphics[width=\linewidth]{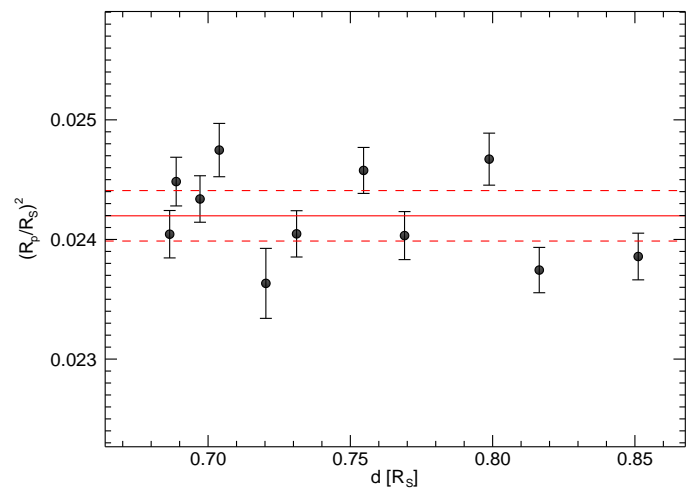}
\caption{Example of the values of ($R_{\rm p}$/$R_{\rm s}$)$^2$ found for one broadband as a function of the distance of the planet from the center of the star ($d$). The red line represents the average ($R_{\rm p}$/$R_{\rm s}$)$^2$
taken as the final value, dashed lines define its uncertainties.}
\label{fig:media}
\end{figure}

We then divide these areas for $A_{\rm ref}$, thus obtaining for each exposure inside the full part of the transit a value for the parameter $\beta$ (see Eq.~\ref{equazbeta}).
Knowing a priori the position of the planet on the stellar disk and the limb darkening coefficients, we divide each $\beta$ for the relative second term of Eq.~\ref{equazbeta}.
For each in-full-transit spectrum we now have an estimate of the ratio ($R_{\rm p}$/$R_{\rm s}$)$^2$ (Fig.~\ref{fig:media}). 
We finally average these values taking as uncertainty their rms. 

It is interesting to note that in few cases the measurements show a clear linear trend as a function of the distance from the center of the star: this is probably due to the fact that the given LD coefficient in use is not completely representative for the star at that moment at that wavelength. In fact the LD coefficients have been demonstrated to be dependent also on the particular situation of the star at the moment of observation \citep[e.g., on the number and size of star spots,][]{csizmadia}. 
In turn, this kind of analysis could be also useful to constrain the LD coefficients of the star.

\begin{figure}[!ht]
\centering
\includegraphics[width=\linewidth]{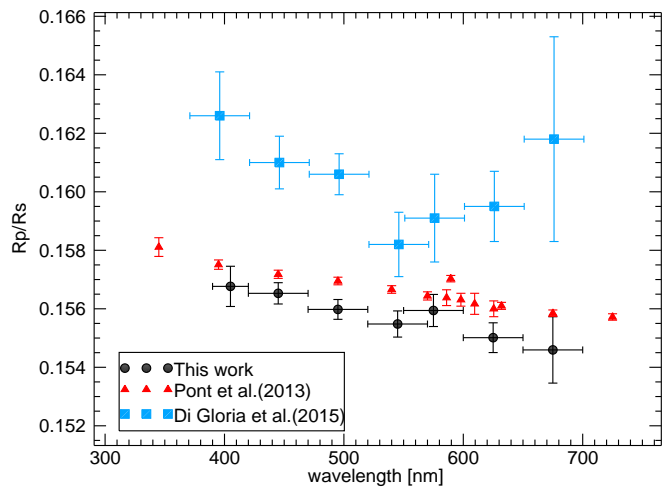}
\caption{
Broadband transmission spectrum of HD~189733b as calculated with this method (black circles).
For comparison, red triangles are HST observations by \citet{pont}, light-blue squares are measurements from \citet{digloria} using the chromatic RM effect on the same dataset of this paper.
The vertical shift is due to the different transit parameters used between different papers.}
\label{fig:plot}
\end{figure}

We repeat the whole procedure for each broadband and for each transit. In this way we obtain one transmission spectrum per transit.
At the end we average the values of the same broadband, obtaining the final average transmission spectrum of HD~189733b (Table~\ref{tabvalori}, Fig.~\ref{fig:plot}).

\begin{table}
\begin{center}
\caption{Values of $R_{\rm p}$/$R_{\rm s}$ for HD~189733b obtained in this work using the chromatic line-profile tomography technique.}
\label{tabvalori}
\footnotesize
\begin{tabular}{ccc}
 \hline\hline
 \noalign{\smallskip}
Group & Passband (nm) & $R_{\rm p}$/$R_{\rm s}$\\
\noalign{\smallskip}
\hline
\noalign{\smallskip}
1 & 400-420  & $0.15677\pm0.00068$\\
2 & 420-470  & $0.15653\pm0.00033$\\
3 & 470-520  & $0.15598\pm0.00030$\\
4 & 520-570  & $0.15548\pm0.00041$\\
5 & 550-600  & $0.15594\pm0.00052$\\
6 & 600-650  & $0.15501\pm0.00048$\\
7 & 650-700  & $0.15460\pm0.00112$\\
\noalign{\smallskip}
 \hline
\end{tabular}
\end{center}
\end{table}

\section{Discussion\label{sec:discussion}}

Our $R_{\rm p}$/$R_{\rm s}$ values derived with chromatic line-profile tomography are in good agreement with those reported by other analyses (Fig.~\ref{fig:plot}).
There is a decreasing in $R_{\rm p}$/$R_{\rm s}$ going toward longer wavelengths that well matches the slope of HST data of \citet[][]{pont}. We did not take into account their correction for star spots since it is well inside our error-bars.
Our uncertainties are of course larger than the ones obtained with HST, but half-size with respect to those obtained by \citet[][]{digloria} who analyzed the same dataset using a different analysis method.
The uncertainties could be reduced by increasing the number of observed transits: with ground-based observations, this should be easier than from space.

\subsection{Single transit events comparison\label{sec:singletransits}}

\begin{figure}[!h]
\centering
\includegraphics[width=\linewidth]{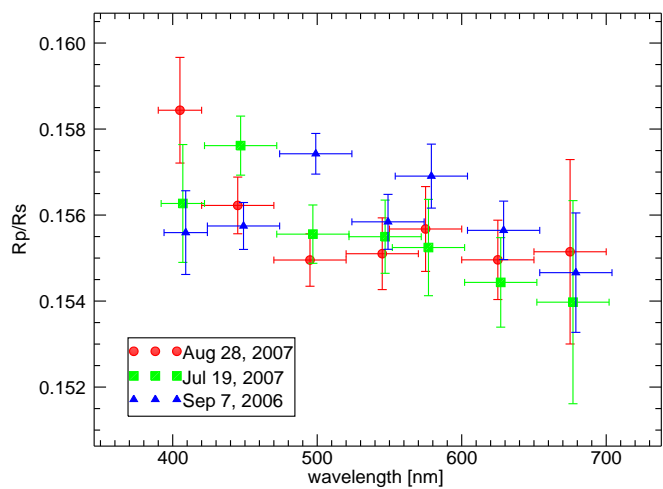}
\caption{
Broadband transmission spectroscopy of HD~189733b calculated for the single transit events recorded with HARPS. 
Red circles refer to the transit of Aug 28th, 2007; green squares to Jul 19th, 2007; blue triangles to Sep 07th, 2006.
The wavelengths for the different transits are slightly shifted for graphic clarity.}
\label{fig:variazioni}
\end{figure}

As mentioned before, the obtained transmission spectrum does not account for any kind of variation due to occulted and unocculted star spots. 
These can affect the CCF, modifying its shape in the same way as the transit of a planet does. As it happens in a transit, star spots occulted by the planet during the transit can bring to an underestimation of the planetary radius, while unocculted ones can bring to an overestimation: this effect is chromatic, and will be thus different depending on the wavelength. 
For this system, this effect has been largely analyzed, with different authors disputing the slope seen in the transmission spectrum
as due to Rayleigh scattering, to stellar activity or both \citep[][]{singSTIS,pont,oshagh,McCullough}.
The detailed study of stellar activity is beyond the scope of the paper, but it is interesting to look at the results obtained in the independent analysis of the three different transits (Fig.~\ref{fig:variazioni}).
While the results at $\lambda>500$ nm are in agreement, we note high variability in the blue part of the spectrum ($\lambda<500$ nm).
Given the high stability demonstrated by HARPS \citep[e.g.,][]{lovis}, 
on the basis of our analysis we stress that this variability could be due to different causes:

\begin{itemize}

\item terrestrial atmosphere: variation in the observing conditions from night to night (e.g., variable seeing, weather) could induce subtle effects on the CCFs \citep[e.g.,][]{cosentino2014}, thus influencing the results;

\item instrumental systematics: in the zone of lower S/N, due to both the spectral energy distribution of the star and the lower efficiency of the instrument, possible systematics could arise between different nights in the CCFs computation;

\item exoweather variability: the data are taken at different times, so in principle variations of planetary radius could be attributed to intrinsic variations in the atmosphere of the planet. This hypothesis is however unlikely;


\item stellar activity: occulted and unocculted star spots affect this kind of measurements in a wavelength-dependent way. Their amount on the stellar surface is obviously changing with time, so their influence on the transmission spectrum can vary for each transit.

\end{itemize}

The same variability going toward blue wavelengths, when looking at the transmission spectrum of HD~189733b taken in different transits, was evident also in \citet[][]{singSTIS} using space-based observations.
This makes us think that the variation observed between different transits is actually due to stellar activity.
We note that HD~189733$\equiv$V452 Vul was known as a rotational variable well before the discovery of the transiting exoplanet. Microvariability ( $\leq0.03$ mag) was discovered by {\it Hipparcos}, while \ion{Ca}{ii} H{\&}K lines were observed in absorption with no variability \citep[][]{cutispoto} and in emission \citep[][]{strassmeier}.

\subsection{A look at sodium\label{sec:sodium}}

The sodium D-lines (588.995 nm and 589.592 nm) have been
already observed in the atmosphere of HD~189733b, in particular using the same dataset as used in this paper \citep[e.g.,][]{wyttenbach}. 
Looking at Fig.~\ref{fig:plot}, it is possible to note in our results a slight deviation from the decreasing radius toward longer wavelengths for the point relative to group 5, that matches in wavelength the variation given by the sodium doublet evident in the measurements from \citet{pont}. 
We stress however that this deviation could not be directly due to sodium: the sodium D-lines are in fact absent in the HARPS line masks and thus they do not contribute to the CCFs and to our analysis.
In our opinion, it is also unlikely that the extended wings of planetary sodium absorption could play a role in this slight increase.


%

\section{Conclusions\label{sec:conclusions}}

We tested a new method for retrieving the broadband transmission spectrum of transiting exoplanets, proving that
chromatic line-profile tomography during transit is a useful tool to provide insights on their atmospheres.
We applied our method to existing high-quality high-resolution transit spectroscopic observations of HD~189733b, together with HD~209458b the most well studied exoplanet so far.
The analysis of three transits observed with HARPS results in an averaged transmission spectrum that is in good agreement with the spectrum obtained from space \citep{pont}, confirming that ground-based high-resolution spectrographs can be used to investigate the broadband transmission spectra of exoplanets.

Comparing one by one the independent results from the three transits analyzed, we can see a disagreement increasing toward the blue part of the spectrum. 
 Even if we cannot completely exclude systematics,
the most probable explanation is that stellar activity plays a role when extracting transmission spectra of exoplanets: care has thus to be taken when claiming for Rayleigh scattering in the atmosphere of planets transiting active stars using only one transit.
Our technique matches the requirement to perform several observations at the same wavelength to quantify the effects of the stellar activity \citep{oshagh}, with the advantage to use the entire optical spectrum to cross-check the results. In such a way it is a valid alternative to multicolor photometry.
With respect to the use of chromatic RM \citep{digloria}, chromatic line-profile tomography has the advantage that it will be useful even in case of planets transiting high $V_{\rm eq}\sin{i}$ stars, for which it is difficult to estimate precise RV values.
The creation of a custom mask for the CCF estimation \citep[e.g., ][]{borsa} will help in further increasing the precision of the results.

Future high-resolution spectrographs like ESPRESSO \citep{espresso} will provide a gain in the S/N, allowing us to use this method also on fainter targets. 
Observing as more planetary atmospheres as possible is needed for a global understanding of atmospheric properties, and any new technique 
can be useful for this scope.


%

\begin{acknowledgements}
We thank the referee for the careful reading of the manuscript and for the interesting checks suggested. 
Based on observations made with ESO Telescopes at the La Silla Paranal
Observatory under programme IDs 072.C-0488(E), 079.C-0127(A) and 079.C-0828(A).
FB acknowledges financial support from INAF through the ''Progetti Premiali'' funding scheme of
the Italian Ministry of Education, University, and Research. 
MR acknowledges financial support from EU FP7 Collaborative Project SPACEINN: {\it Exploitation of Space Data for Innovative Helio- and Asteroseismology}.
\end{acknowledgements}

%

\end{document}